\documentclass[showkeys,reprint,
 amsmath,amssymb,
 aps,
pra,
]{revtex4-2}

\usepackage[margin=1.5cm]{geometry}
\usepackage[all]{xy}
\usepackage{amssymb,mathtools,hyperref} 
\usepackage{amsmath} 
\usepackage{graphicx} 
\usepackage{ifthen}
\usepackage{pgfmath}
\usepackage[usenames, dvipsnames]{xcolor}
\usepackage{tikz}
\usetikzlibrary{calc,arrows}  
\usepackage{enumerate,comment}
\usepackage{subcaption}
\usepackage[font=small,labelfont=bf,
   justification=justified,
   format=plain]{caption} 

\usepackage{lipsum,xcolor,comment}
\usepackage{amsmath,amssymb,stmaryrd,amsthm}

\theoremstyle{definition}

\newcommand{\G}{\mathsf{G}}
\renewcommand{\H}{\mathsf{H}}
\newcommand{\M}{\mathsf{M}}

\newcommand{\U}{\mathsf{U}}
\newcommand{\dU}{\dot{\mathsf{U}}}
\renewcommand{\u}{\mathsf{u}}

\newcommand{\A}{\mathsf{A}}
\newcommand{\B}{\mathsf{B}}

\newcommand{\R}{\mathsf{R}}

\usepackage{dcolumn,graphicx}
\usepackage{bm}

\usepackage{afterpage}

\begin{document}

\title{Novel frame changes for quantum physics}

\author{Pierre-Louis Giscard$^\dagger$}
 \email{giscard@univ-littoral.fr}

 \author{Omid Faizy$^{\dagger,\ast}$}
 \author{Christian Bonhomme$^{\ast}$}
 
\affiliation{$^\dagger$Laboratoire de Mathématiques Pures et Appliquées Joseph Liouville, Université du Littoral Côte d'Opale, 50 rue Ferdinand Buisson, CS 80699, 62228 Calais, France.}%

\affiliation{$^\ast$Laboratoire de Chimie de la Matière Condensée de Paris, UMR CNRS 7574, Sorbonne Université, 4, place Jussieu, 75252 Paris, France.}

\date{\today}

\begin{abstract}
We present novel, exotic types of frame changes for the calculation of quantum evolution operators. 
We detail in particular the biframe, in which a physical system's evolution is seen in an equal mixture of two different standard frames at once. We prove that, in the biframe, convergence of all series expansions of the solution is quadratically faster than in `conventional' frames. That is, if in laboratory frame or after a standard frame change the error at order $n$ of some perturbative series expansion of the evolution operator is on the order of $\epsilon^n$, $0<\epsilon<1$, for a computational cost $C(n)$ then it is on the order of $\epsilon^{2n+1}$ in the biframe for the same computational cost. We demonstrate that biframe is one of an infinite family of novel frames, some of which lead to higher accelerations but require more computations to set up initially, leading to a trade-off between acceleration and computational burden.
\end{abstract}

\maketitle

	%\date{\today}

 \section{Introduction}
In this work we consider physical systems whose properties or dynamics is described by a system of ordinary linear differential equations with possibly time-dependent coefficients, 
\begin{equation}\label{NODEPhys}
    \frac{d}{dt}\U(t)=\A(t)\U(t),\quad \U(0)=\mathsf{Id}.
\end{equation}
Chief examples include closed quantum systems obeying Schr\"{o}dinger equation with $\A(t)=-i \H(t)$, $\H(t)$ being the quantum Hamiltonian; as well as the partition function $\mathcal{Z}(\beta)$ of statistical physics, which being the exponential of the Hamiltonian operator $\exp(-\beta \mathcal{H})$, solves the ordinary differential equation $\dot{\mathcal{Z}}(\beta)=-\mathcal{H}\,\mathcal{Z}(\beta)$. 

Calculating analytically the solution $\U$ (called evolution operator) of Eq.~(\ref{NODEPhys}) when $\A(t)$ does not commute with itself at different times is notoriously difficult. Formally, the solution is designated as a time-ordered exponential, 
\begin{align*}
    \U(t) &= \mathcal{T}e^{\int_0^t\A(\tau)d\tau}\\
&=\mathsf{Id}+\int_0^t\A(\tau)d\tau+\int_0^t\int_{0}^{\tau_1}\A(\tau_1)\A(\tau_2)d\tau_2 d\tau_1+\cdots
\end{align*}
with $\mathcal{T}$ the time-ordering operator.
The lack of self-commutativity $\A(t)\A(s)\neq \A(s)\A(t)$ hinders simplications in the Dyson series above and leads to a break down in the invariance of the evolution operator under time translations $\U(t,s)\neq \U(t-s,0)$. That is, evolving the solution from a time $s$ to a time $t>s$ is different from evolving it from 0 to $t-s$. This critical observation indicates that the correct general mathematical framework to determine $\U$ is that of bivariate functions and matrices, i.e. depending on two time-variables, even though only $\U(t):=\U(t,0)$ is usually desired in physics. Volterra and P\'er\`es were the first to consider the algebraic structures necessitated by this two-times approach to differential equations in a pioneering study completed in the early 1920s \cite{Volterra1924}. Lacking a proper understanding of distributions at the time, their mathematics encountered many profound obstacles and their work was subsequently overlooked. In a series of recent developments the operation Volterra introduced \cite{Volterra1911} to deal with iterated integrals (the Volterra composition) has been given its proper context and generalization within the theory of distributions, yielding the $\star$ product, 
\begin{equation}\label{fgStar}
(f\star g)(t,s):=\int_{-\infty}^{+\infty} f(t,\tau)g(\tau,s)d\tau. 
\end{equation}
Here $f,g\in\mathcal{D}$ belong to a certain set of distributions which we do not need to explicit, see \cite{Ryckebusch2023} for a rigorous presentation and Appendix~\ref{ProperStar}. The $\star$ product extends naturally to matrices in $\mathsf{A},\B\in\mathcal{D}^{N\times N}$ with
\begin{equation}\label{ABStar}
\big(\mathsf{A}\star \B\big)(t,s) =\int_{-\infty}^{+\infty} \mathsf{A}(t,\tau).\B(\tau,s)d\tau.
\end{equation}
This product has a unit, $\mathsf{I}_\star:=\mathsf{Id}\, \delta(t-s)$.
For our purposes here the most important fruit of this formalism is that even if $\A(t)$ does not commute with itself at all times the differential system of Eq.~(\ref{NODEPhys}) has Green's function 
\begin{equation}\label{GResol}
 \G(t,s) = \big(\mathsf{I}_\star- \A(t) \Theta(t-s) \big)^{\star -1},
\end{equation}
and $\U(t,s)=\int_s^t \G(\tau,s) d\tau$ is the integral of $\G$. In this expression, $\Theta(t-s)$ is the Heaviside theta function, $\Theta(x)=1$ if $x\geq 0$ and 0 otherwise. Interestingly, it arises out of mathematical necessity but encodes a physical reality: causality, $s$ being the time at which initial conditions are imposed and the drive by $\A$ starts, while $t$ is the time of observation. The crucial observation is that $\G$ is a true resolvent of $\A$: this means that it is amenable to all techniques from ordinary linear algebra, so long as $\star$ products are substituted in place of ordinary matrix products. 

An implication of this observation concerns frame changes in physics. By frame change, we here mean in a narrow sense those implemented mathematically on differential systems like Eq.~(\ref{NODEPhys}). Concretely it is well known that the evolution operator $\U$ solution of Eq.~(\ref{NODEPhys}) can be recast as
$$
\U(t) = \U_\B(t) \mathcal{T} e^{\int_0^t \U_\B^\dagger(\tau) \A(\tau) \mathsf{U}_\B(\tau) d\tau},
$$
with, for discrete systems, $\B(t)$ a matrix of the same size as $\A$ and $\U_\B$ the evolution operator solution of $\dot{\U}_\B=\B\,.\U$. This corresponds to seeing the dynamics driven by $\A$ from a new frame whose movement is due to $\B$.
We shall see that this technique is in fact the simplest of infinitely many novel types of frame changes which offer progressively higher computational rewards as they become increasingly intricate. 

\section{Frame changes as linear $\star$-algebra}
\subsection{Ordinary frame change}\label{StdFrame}
Instead of directly aiming for frame changes, let us consider the following problem: suppose that the matrix $\A(t)$ can be expressed as the sum of two parts $$\A(t)=\A_0(t)+\A_1(t)$$ and that one wishes to relate the evolution operator $\U$ solution of $\dot{\U}=\A(t)\U$ to the evolution operators $\U_0$ and $\U_1$ solving the equations $\dot{\U}_0=\A_0(t)\U_0$ and $\dot{\U}_1=\A_1(t)\U_1$, respectively. 
Given that the Green's function $\G$ is a $\star$-resolvent as shown by Eq.~(\ref{GResol}), the question is the same as expressing, for two matrices $\M_0$ and $\M_1$, the quantity $\mathsf{R}:=(\mathsf{I}-\M_0- \M_1)^{-1}$ in terms of the `isolated' resolvents $\mathsf{R}_0:=(\mathsf{I}-\M_0)^{-1}$ and $\mathsf{R}_1:=(\mathsf{I}-\M_1)^{-1}$. 
Standard linear algebra provides many such representations. For example,
\begin{subequations}\label{FrameChangeRes1}
\begin{align}
\frac{1}{\mathsf{I}-\M_0-\M_1}&=\frac{1}{\mathsf{I}-\M_1}\frac{1}{\mathsf{I}-\M_0\frac{1}{\mathsf{I}-\M_1}},\\
\intertext{which we wrote in fraction form for improved readability but which should be understood as the statement that }
\mathsf{R}&=\mathsf{R}_1.\left(\mathsf{I}-\M_0.\mathsf{R}_1\right)^{-1}.\label{SimpleExp}
\end{align}
\end{subequations}
This equation remains valid should $\M_0$ and $\M_1$ not commute. 

Let us see what this implies for Green's functions. Turning back to $\A(t)=\A_0(t)+\A_1(t)$, let $\mathsf{G}_{1}=(\mathsf{I}_\star -\A_1\Theta)^{\star-1}$ designate the Green's function associated to $\A_1$. Then Eq.~(\ref{SimpleExp}) is equivalent to the statement that
$$
\G = \G_1 \star \left(\mathsf{I}_\star -\A_0\Theta \star \G_1\right)^{\star -1},
$$
and the full evolution operator $\U(t)$ obeys,
\begin{equation}\label{Ustdframe}
\U = \U_1 \star \left(\mathsf{I}_\star - \A_0\Theta \star \G_1\right)^{\star -1}.
\end{equation}
Evaluating the $\star$-products and inverses explicitly in the above yields (Appendix~\ref{EvalFrameStd}), 
\begin{subequations}
\begin{align}\label{StdExplicit}
\mathcal{F}(t,s)&=\int_s^t \U_1^{-1}(\tau)\A_0(\tau)\U_1(\tau)d\tau,\\
\U(t,s)&=\U_1(t)\,\mathcal{T}e^{\mathcal{F}(t,s) }\,\U_1^{-1}(s),
\end{align}
\end{subequations}
which is the standard frame change formula. This proof of the formula is neither better nor simpler than the classical proofs relying on differential calculus, but the point is we obtained it from the purely linear algebraic statement Eq.~(\ref{SimpleExp}). This suggests that changing frame is one of a very large reservoir of differential transformations which are $\star$-linear algebraic results in disguise. Indeed, plenty more such formulas must exist since resolvents of ordinary matrices satisfy a host of relations in the spirit of Eq.~(\ref{SimpleExp}), all of which ought to correspond to valid results concerning Green's functions and evolution operators. 

We put this reasoning to the test by looking for an hitherto unknown types of frame change, in particular one that does not break the symmetric roles played by $\A_0$ and $\A_1$ in $\A$: the biframe.
As an additional bonus, we will show that contrary to the standard frame change, advanced frame changes such as the biframe intrinsically accelerate the convergence of perturbation series. 

\subsection{Biframe change}
Coming back to Eq.~(\ref{SimpleExp}), we see that the formula treats $\M_0$ and $\M_1$ differently. One might instead wish for a more symmetrical treatment of both parts that does not give undue importance to one over the other. This is achieved by the following formula (which is certainly not the only possible solution to the quest for symmetry!), which is just as easily verified using first year calculus rules:
\begin{subequations}\label{FrameChangeRes2}
\begin{align}
\frac{1}{\mathsf{I}-\M_0-\M_1}&=\frac{1}{\mathsf{I}-\M_0}\frac{1}{\mathsf{I}-\frac{\M_1}{\mathsf{I}-\M_1}\frac{\M_0}{\mathsf{I}-\M_0}}\frac{1}{\mathsf{I}-\M_1},\\
\intertext{more rigorously,}
\mathsf{R}&=\mathsf{R}_0.\big(\mathsf{I}-\M_1.\mathsf{R}_1.\M_0.\mathsf{R}_0\big)^{-1}.\mathsf{R}_1.\label{RRRresult}
\end{align}
\end{subequations}
Again this remains true when $\M_0$ and $\M_1$ do not commute. 

Let us now see what this implies for the evolution operator $\U$. Taking $\A(t)=\A_0(t)+\A_1(t)$, replacing matrix products by $\star$-products and ordinary resolvents $\mathsf{R}$ by Green's functions one gets,
\begin{align}
\G&=\G_0\star\big(\mathsf{I}_\star -  \A_1\Theta\star\G_1\star \A_0\Theta \star\G_0\big)^{\star-1}\star \G_1,\label{basicG}
\intertext{and from there}
\U&=\U_0\star\big(\mathsf{I}_\star -  \A_1\Theta\star \G_1\star \A_0\Theta \star\G_0\big)^{\star-1}\star \G_1.\label{basicU}
\end{align}
In spite of appearances with e.g. $\A_1$ on the left of $\A_0$ inside the $\star$-resolvent, both parts play exactly equivalent roles in these results. For example, noting that $\dot{\U}_i=\A_{i}\Theta\star \G_i$, Eqs.~(\ref{basicG}, \ref{basicU}) are equivalent to the following result for the derivative $\dU$ (Appendix~\ref{AppendixAlternative}),
\begin{align*}
\dU &=\dU_0+\dU_1+\dU_0\star \dU_1+\dU_1\star\dU_0 +\dU_0\star \dU_1\star \dU_0\\&\hspace{5mm}+\dU_1\star \dU_0\star \dU_1 +\dU_0\star\dU_1\star \dU_0\star \dU_1+\cdots,
\end{align*}
which is manifestly invariant under the exchange of indices $0\leftrightarrow 1$.

Evaluating the $\star$-products in the  $\star$-resolvent and keeping in mind that such a resolvent is a time-ordered exponential, Eq.~(\ref{basicU}) is equivalent to
\begin{subequations}\label{BiFrame1}
\begin{align}
&\mathcal{B}(t,s):=\A_1(t)\U_1(t)\int_s^t\U^{-1}_1(\tau)\A_0(\tau)\U_0(\tau)d\tau\,\U_0^{-1}(s),\label{BiframeO}\\
&\U=\U_0\star  \mathcal{T}e^{\mathcal{B}(t,s)}\star \G_1.\label{BlueVersion}
\end{align}
\end{subequations}
This is the biframe change. A detailed proof of this result is presented in Appendix~\ref{ProofBlue}. An alternative though completely equivalent form of the biframe is obtained on re-arranging the terms in the formula above, yielding (proof in Appendix~\ref{ProofsDetail}), 
\begin{subequations}
\begin{align}
&\mathcal{B}_2(t,s):=\A_1(t)\U_0(t)\int_s^t\U^{-1}_0(\tau)\A_0(\tau)\U_1(\tau)d\tau\,\U_1^{-1}(s),\label{BiframeO2}\\
&\U=\U_0\star \G_1\star \mathcal{T}e^{\mathcal{B}_2(t,s)}\label{RedVersionmain}.
\end{align}
\end{subequations}
In Appendix~\ref{BiframeConstant} we present simplifications of the above results should $\A$ be a constant matrix.
Before we see how to implement the biframe in a concrete example, we show that it is worth the effort.

\subsection{Intrinsic acceleration of perturbation series}\label{BonusAcc}
While standard frame changes can yield structurally simpler Hamiltonians, the frame change itself is not expected to inherently speed up perturbation series. To see this, coming back to Section \ref{StdFrame}, we showed that the standard frame change is equivalent to the $\star$-linear algebraic statement Eq.~(\ref{Ustdframe}). Expanding this expression as a Dyson series shows that the $m$th order approximation $\U_{\text{std-frame}}^{[m]}$ of $\U$  after a standard frame change defined as 
$$
\U_{\text{std-frame}}^{[m]}:=\U_1\star\sum_{k=0}^m (\A_0\Theta \star \G_1)^{\star k},
$$ 
correctly reproduces up to order $m$ of the original Dyson series expansion of $\U$ in the laboratory frame. In other terms, the change of frame did not alter the approximation order $m$ and therefore did not accelerate the series expansion of $\U$. 

In contrast, the biframe change intrinsically produces an acceleration of this expansion. To see why, let us go back to the standard linear algebra. Suppose for this discussion that some matrix $\mathsf{M}$ has a spectral radius $\rho(\mathsf{M})<1$ so that the Neumann series $\sum_{k=0}^\infty \mathsf{M}^k$ converges. This series is a natural perturbative expansion scheme (with a small parameter $\rho(\M)$) for the resolvent $\R:=(\mathsf{I}-\mathsf{M})^{-1}$ . Since $\mathsf{M}^{k+1}=\mathsf{M}.\mathsf{M}^k$, then at order $m>1$ the standard approximation $\R_{\text{std}}^{[m]}:=\sum_{k=0}^m \mathsf{M}^k$ can be calculated with $m-1$ matrix products. 
At the same time, it is equally true that
$$
\R=(\mathsf{I}+\mathsf{M}).(\mathsf{I}-\M^2)^{-1},
$$
so that another expansion scheme for the resolvent is
$\R^{[m]}:=(\mathsf{I}+\mathsf{M}).\sum_{k=0}^m \mathsf{N}^k$, with $\mathsf{N}=\mathsf{M}^2$. Calculating this requires $(m-1)+1+1=m+1$ matrix products, $m-1$ of which come from  $\sum_{k=0}^m \mathsf{N}^k$, one more from calculating $\mathsf{N}=\M^2$ and another one for the final multiplication by $(\mathsf{I}+\M)$. Yet, expanding $\R^{[m]}$ immediately shows that it reproduces up to order $2m+1$ of the original Neumann series $\R^{[m]}=\R^{[2m+1]}_\text{std}$, roughly doubling the approximation order at the same computational cost. 

Eq.~(\ref{RRRresult}) is a peculiar instance of the above construction. Observe that for $k\geq0$, the expression
$
\R_0 \big(\M_1\R_1\M_0\R_0\big)^k \R_1
$
comprises $2k+1$ exchanges between the 0 and 1 indices, precisely as much as $(\M^2)^n=(\M_0+\M_1)^k$. Since furthermore $\R_i:=(\mathsf{I}-\M_i)^{-1}$, $i=0,1$, then the truncated series
$$
\R^{[m]}:=\R_0 .\,\sum_{k=0}^m\big(\M_1.\R_1.\M_0.\R_0\big)^{k}\,. \R_1,
$$
is verified to reproduce the first $2m+1$ terms of the standard Neumann series while necessitating only $m+1$ matrix products.\\

Given that solving N-ODEs is nothing but doing $\star$-linear algebra, these results hold true for Green's functions. Now the biframe operator $\mathcal{B}$  of Eq.~(\ref{BiframeO}) involves the two parts $\A_1$ and $\A_0$ so that the Dyson series of its time-ordered exponential, if truncated at order $m$, necessitates $m+1$ $\star$-products (i.e. iterated integrals with matrix products) while yielding an expression that is exact up to order $2m+1$ of the Dyson series expansion of $\U$ after a standard frame change. Denoting by $\epsilon^n$ the error at order $n$ of the Dyson expansion without frame change (or equally after a standard frame change), the error at the same order of expansion in the biframe is thus $\epsilon^{2n+1}$, quadratically better at nearly fixed computational cost. From an information point of view, this is because we assume that we exactly know the Green's functions of the isolated parts $\G_0$ and $\G_1$, the biframe change leveraging this knowledge into an acceleration of the original series.

\subsection{Further frame changes}
Given the new perspective on frame changes as $\star$-linear algebraic statements on resolvents (like Eqs.~\ref{FrameChangeRes1}, \ref{FrameChangeRes2}), there exists as many frame changes as there are ways to express the resolvent $\R$ of a matrix $\M:=\sum_i \M_i$ in terms of its $\M_i$ parts and their the resolvents. Many of these exotic frame changes can produce further accelerations depending on the known quantities or symmetries that one wishes to preserve. For instance, let us devise a triframe change where the systems is seen in three different frames at once, each generated by one part of an operator $\A(t)=\A_0(t)+\A_1(t)+\A_2(t)$. This is equivalent to asking for the ordinary resolvent $\R$ of a matrix $\M=\M_0+\M_1+\M_2$ in terms of the resolvents $\R_{i}$, $i=0,1,2$ in a way that these play equivalent roles. Standard linear algebra indicates that,
\begin{align*}
\R&=\R_0.(\mathsf{I}-\M_1.\R_1.\M_0.\R_0)^{-1}.\R_1\\
&~~~~~.\Big(\mathsf{I}-\M_2.\R_2.\left(\M_0.\R_0.(\mathsf{I}-\M_1.\R_1.\M_0.\R_0)^{-1}.\R_1\right.\\
&\left.\hspace{15mm}+\M_1.\R_1.(\mathsf{I}-\M_0.\R_0.\M_1.\R_1)^{-1}.\R_0\right)\Big)^{-
   1}.\R_2,
\end{align*}
For evolution operators, this is therefore
\begin{align*}
\U&= \U_0\star(\mathsf{I}_\star-\dU_1\star\dU_0)^{-1}\star\G_1\\
&~~~~~\star\left(\mathsf{I}_\star-\dU_2\star\left(\dU_0\star(\mathsf{I}_\star-\dU_1\star\dU_0)^{\star-1}\star\G_1\right.\right.\\
&\left.\left.\hspace{15mm}+~\dU_1\star(\mathsf{I}_\star-\dU_0\star\dU_1)^{\star-1}\star\G_0\right)\right)^{\star-1}\!\!\!\star\G_2
\end{align*}
where $\dU_i:=\A_i(t)\Theta\star \G_i$ and $\G_i$ is the Green's function associated to Hamiltonian $\H_i(t)$.
Once expanded as a series the triframe generates the expansion 
\begin{align*}
\dU =&~ \dU_0 + \dU_1+\dU_2+ \dU_0\star \dU_1+\dU_1\star \dU_0+\dU_0\star \dU_2+\dU_2\star \dU_0\\&+\dU_1\star \dU_2+\dU_2\star \dU_1+\dU_0\star \dU_1\star \dU_2 + \dU_0\star \dU_2\star \dU_1+\cdots
\end{align*}
which is invariant under  exchange of indices $0\leftrightarrow 1\leftrightarrow 2$.

To determine the acceleration of series expansions in the triframe we follow the analysis of section \ref{BonusAcc}. Let $\mathsf{L}:=\M^3$, then observe that
$$
\R=(\mathsf{I}+\mathsf{M}+\mathsf{M}^2).(\mathsf{I}-\mathsf{L})^{-1},
$$
which implies that $(\mathsf{I}+\mathsf{M}+\mathsf{M}^2).\sum_{k=0}^m \mathsf{L}^k$  is a series expansion of $\R$ necessitating only $m+2$ matrix products to reach order $3m+2$ of the original Neumann series expansion of $\R$ in powers of $\M$. This implies that cutting the original matrix $\M$ into three parts, each of which, if taken in isolation, has a known Green's function, is leveraged into a cubic acceleration of the Dyson series for $\U$ in the triframe. 
Note that the triframe is equivalent to applying a biframe change twice, which is not particularly creative: many more ways of doing exist, e.g. if only some but not all isolated Green's functions $\G_i$ are known or if one wants to single emphasize the role of one ordinary frame over the others. For example one could choose to mix a standard frame change with a biframe, leading to smaller acceleration of convergence as compared to the triframe but possibly easier to wield operators.

Clearly, frame changes tailored to any one problem can be devised by leveraging known quantities such as Green's functions of isolated parts into a useful expression with profitable acceleration for perturbative expansions.

\section{Example}

\subsection{Acceleration of Dyson series for a time-dependent Hamiltonian}
Consider the Schr\"{o}dinger equation with Hamiltonian
$$
\mathsf{H}(t)=\frac{\omega_0}{2}\sigma_z + 2\beta \cos(\omega t)\sigma_x.
$$ 
Denote $\H_0=(\omega_0/2)\sigma_z$, $\H_1=2\beta\cos(\omega t)\sigma_x$. In consequence, there are two natural standard frame changes for this Hamiltonian. The first is, 
\begin{align*}
\H\longrightarrow  \H_\text{std,\,0}(t)&:=\U_0^\dagger(t).\H.\U_0(t),\\
&= \begin{pmatrix} \frac{\omega_0}{2} & 2 \beta  e^{i \omega_0 t} \cos (\omega t ) \\
 2 \beta  e^{-i \omega_0 t} \cos (\omega t ) & -\frac{\omega_0}{2} 
 \end{pmatrix}.
 \end{align*}
This frame change is mostly used to justify the rotating wave-approximation in near-resonant cases $\omega_0\approx \omega$ with both large.
The second frame change leads to
\begin{align*}
\H\longrightarrow \, &\H_\text{std,\,1}(t):=\U_1^\dagger(t).\H.\U_1(t),\\
&=2\beta\cos(\omega t)\sigma_x+\frac{\omega_0}{2}\sin\left(\frac{4 \beta}{\omega }  \sin (\omega t )\right)\sigma_y\\&\hspace{5mm}+\frac{\omega_0}{2}\cos\left(\frac{4 \beta}{\omega }  \sin (\omega t )\right)\sigma_z
 \end{align*}
 This choice is advantageous in the situation where $\omega\gg1$, the time-average of $ \H_\text{std,\,1}$ being the first order of the high-frequency expansion of $\mathsf{U}(t)$ \cite{Mikami2016, Venkatraman20222}. Subsequent orders correspond to higher moments of  $\H_\text{std,\,1}$. We now turn to the biframe, the driving operator is
\begin{align*}
&\mathcal{B}(t,s)=\\
&\beta  \omega_0 \cos (\omega t) \cos \left(\frac{2 \beta}{\omega}  \sin (\omega t)\right) \begin{pmatrix}
    -i S(t,s) & \overline{C}(t,s) \\
 - C(t,s) & i \overline{S}(t,s) 
\end{pmatrix}\\
&+\beta  \omega_0 \cos (\omega t)  \sin \left(\frac{2 \beta}{\omega}  \sin (\omega t)\right)\begin{pmatrix}
i  C(t,s)  & \overline{S}(t,s)  \\
 - S(t,s)  & -i \overline{C}(t,s)  
\end{pmatrix},
\end{align*}
with 
\begin{align*}
S(t,s):= e^{\frac{1}{2}i \omega_0 s}\int_s^t e^{-\frac{1}{2} i \omega_0\tau} \sin \left(\frac{2 \beta}{\omega
   }  \sin (\omega\tau )\right)d\tau,\\
   C(t,s):=e^{\frac{1}{2} i \omega_0 s } \int_s^t e^{-\frac{1}{2} i \omega_0\tau} \cos \left(\frac{2 \beta}{\omega}  \sin (\omega\tau )\right)d\tau,
\end{align*}
and $\overline{S}$, $\overline{C}$ are the conjugates of $S$ and of $C$, respectively. 
In order to demonstrate the acceleration of perturbative expansions that is intrinsic to the biframe, we present a numerical comparison of the Dyson expansion in the laboratory frame (no frame change), after the two standard frame changes and in the biframe. To that end, define the $m$th-order Dyson approximations 
\begin{align*}
&\U_\text{lab}^{[m]}:=\Theta\star \sum_{k=0}^m \H^{\star k},\\
&\U_\text{std-frame}^{[m]}:=\U_1\star \sum_{k=0}^m \H_{\text{std}}^{\star k},\\
&\U_\text{biframe}^{[m]}:=\U_0\star \sum_{k=0}^m (\mathcal{B}\Theta)^{\star k} \star \G_1,
\end{align*}
where $\H_\text{std}$ is the Hamiltonian after a standard frame change, i.e. either $\H_\text{std,\,0}$ or $\H_\text{std,\,1}$. 
Recall that for any matrix comprising smooth functions of time, $(\A\Theta)^{\star k}$ reduces to the $k$th iterated integral of $\A(t)$ so that $\sum_k (\A\Theta)^{\star k}$ is the Dyson series $\mathsf{I}\delta(t-s) + \A(t)+\int_s^t \A(t)\A(\tau)d\tau +\cdots$ as claimed.

\begin{figure}[t!]  
     \centering
\includegraphics[width=.47\textwidth]{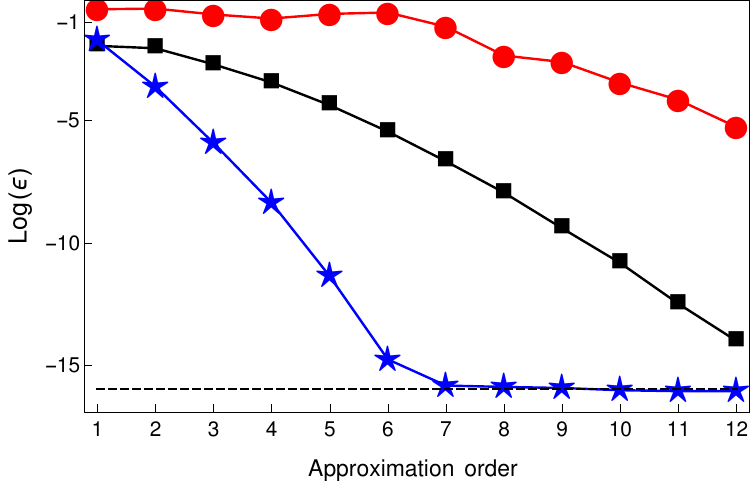}
\vspace{-1mm}
    \caption{Logarithm of the relative error $\epsilon$ of Eq.~(\ref{eq:froerr}) for the first 12 orders of the Dyson series expansion in the laboratory frame (red circles and line), after a standard frame change (black squares and line) and in the biframe (blue stars and line). Scaled parameters: $\omega_0/\omega\simeq 0.67$, $\beta/\omega\simeq 0.53$, total scaled simulation time $\omega T=6$. The dashed straight line at the bottom represents machine precision. The poor performance of the laboratory frame
    is due to the inability of its low orders to correctly fit the behavior at long times $t\sim T$, which heavily degrades the overall error measure $\epsilon$.
    }
    \label{ErrorFrame}
\end{figure}

We quantify the error associated with each order $m$ approximation as follows.
Let $\U_{r}$ be the reference evolution operator as evaluated within machine precision (relative and absolute tolerances set to $10^{-16}$ for each entry) by a standard numerical solver (in this case Mathematica's NDSolve).
We evaluate the accuracy of $\U_{\text{lab}}^{[m]}$, $\U_{\text{std-frame}}^{[m]}$ and  $\U_{\text{biframe}}^{[m]}$ by evaluating the deviation from 1 of their normalised Frobenius scalar products with $\U_r$ over a total evolution time $T$,
\begin{equation}
\label{eq:froerr}
\epsilon := \frac{1}{T}\int_{0}^T 1-\frac{\mathrm{Tr}\big(\U_r^\dagger(\tau)\U^{[m]}(\tau)\big)}{\sqrt{\|\U_r(\tau)\|_F\,\|\U^{[m]}(\tau)\|_F}}d\tau,
\end{equation}
which is the relative error on $\U^{[m]}$ with respect to the reference solution. 
Here $\|\A\|_F:=\mathrm{Tr}(\A^\dagger \A)$ designates the Frobenius norm of matrix $\A$. As constructed above, the relative error $\epsilon$ evaluates to 0 if and only if $\U^{[m]}(t)=\U(t)$ exactly at all times between $t=0$ and $t=T$. In Fig.~\ref{ErrorFrame} we show the relative errors in all three frames and up to order $m=12$ of the Dyson series. Numerical computations of $\star$-products and $\star$-resolvents rely on trapezoidal quadrature as detailed in \cite{DD21}. 

\section{Conclusion}
We show that changing frame in the context of autonomous and non-autonomous systems of coupled ordinary linear differential equations such a Schr\"{o}dinger's, is an instance of a simple linear algebraic statement with respect to a generalized convolution-like product, known as the $\star$-product. Exploiting this observation, we showed that infinitely many novel frames exist and that these intrinsically accelerate perturbative expansions of evolution operators and partition functions. 

The results presented here are certainly not limited to quantum mechanics nor to first order linear differential equations. Indeed, since any $n$th order non-autonomous linear differential equation is equivalent to a first order non-autonomous $n\times n$ system of coupled linear differential equations, then it is amenable to all the frame changes presented in this work. We shall illustrate this in a coming work with novel perturbative expansion of general Heun functions around  hypergeometric solutions of Heun equations.

\subsection*{Acknowledgements}
 This work is supported by the French National Research Agency ANR-20-CE29-0007 project \textsc{Magica}.\\

 \appendix

\section{Properties of the $\star$ product}\label{ProperStar}
\subsection{Relation to Volterra composition}
For bivariate functions $f$ and $g$ that are smooth in both of their arguments $t$ and $s$, the $\star$-product between $f(t,s)\Theta(t-s)$ and $g(t,s)\Theta(t-s)$ is the Volterra composition of the first kind $\star_v$ of the smooth functions $f$ and $g$,
\begin{align*}
&(f\Theta \star g \Theta)(t,s)\\
&\hspace{3mm}=\int_{-\infty}^{\infty} f(t,\tau)\,\Theta(t-\tau)\,g(\tau,s)\,\Theta(\tau-s)d\tau,\\
&\hspace{3mm}= \int_s^t f(t,\tau) g(\tau,s) d\tau\,\Theta(t-s)=(f\star_v g)(t,s)\, \Theta.
\end{align*}
The Volterra composition, which naturally appears in the Picard iterations is mathematically ill-behaved in general, lacking a proper unit and inverses for example \cite{Volterra1924}. All of these issues are lifted by the $\star$-product which provides the correct distributional context for non-autonomous linear differential systems and beyond (especially fractional and non-linear ordinary differential systems). 

\subsection{Relation to the convolution}
In general both the $\star$-product and the Volterra composition differ from convolutions because $f,g$ in Eq.~(\ref{fgStar}) and $\mathsf{A}, \B$ in Eq.~(\ref{ABStar}) may not depend solely on the difference between their arguments. If we explicitely consider two elements $k,l\in\mathcal{D}$ that are invariant under time translations, $k(t,s) =k(t-s,0)\equiv k(t-s)$ and $l(t,s)=l(t-s,0)\equiv l(t-s)$; then they both \textit{effectively} depend on the single variable $t-s$ and their $\star$-product reduces to a convolution
\begin{align*}
(k\star l)(t,s) &= \int_{-\infty}^{\infty} k(t,\sigma)l(\sigma,s)d\sigma,\\
&\equiv \int_{-\infty}^{\infty} k(t-\sigma)l(\sigma-s)d\sigma=(k\ast l)(t,s) 
\end{align*}
In the context of ODEs this happens if and only if the coefficient matrix $\A$ commutes with itself at all times. Indeed, this is tied to the behavior of the evolution operator $\U$ that solves the N-ODE $\dU=\A\U$. In general, $\A$ does not commute with itself at all times if and only if $\U$ is not invariant under time translations: $\U(t,s)\neq \U(t-s,0)$. This confirms the necessity of working with bivariate objects in the most general theory. 

Note that in all cases, the usual one-variable evolution operator is related to the bivariate one by the semi-group property, as follows $\U(t,s)=\U(t)\U^{-1}(s)$. In an abuse of notation we will denote the bivariate and one-variable evolution operators by the same letter $\U$.

\subsection{Solution of non-autonomous systems with the $\star$-formalism}
Let $\A(t)$ be a time-dependent matrix all of whose entries are smooth functions of time and consider the system of linear Non-autonomous Ordinary Differential Equations (N-ODE) 
\begin{equation}\label{NODE}
\dU(t,s)=\A(t)\U(t,s),\quad \U(s,s)=\mathsf{Id},
\end{equation}
for $t\geq s$ two real variables. 
Here $\U(t,s)$ designates the matrix solution of the system, termed its evolution operator. 
The $\star$-product turns such N-ODEs and their solutions into purely linear $\star$-algebra.
To see this let $\U(t,s):=\U(t,s)\Theta(t-s)\in\mathcal{D}$ and introduce the Green's function $\G:=(\delta'\mathsf{Id}) \star \U  = \mathsf{Id}_\star + \dU\Theta$. Then the N-ODE Eq.~(\ref{NODE}) becomes \cite{Giscard2015}
\begin{equation}\label{GreenEq1}
\G-\mathsf{Id}_\star =\A\Theta \star \G.
\end{equation} 
The solution of Eq.~(\ref{GreenEq1}) is 
\begin{equation}
 \G = \big(\mathsf{Id}_\star - \A\Theta\big)^{\star -1},
\end{equation}
 and $\U = \Theta \star \G$ (because $\Theta\star \delta'=\mathsf{Id}_\star$). Here the existence of the $\star$-inverse in $\G$ is guaranteed by standard results from the analysis of Picard iterations \cite{Giscard2015} since 
 \begin{align*}
 \G &= \sum_{n\geq 0} (\A\Theta)^{\star n},\\
 &=\mathsf{I}_\star + \A\Theta + \int_s^t \A(t)\A(\tau) d\tau \,\Theta\,+ \\
 &\hspace{8mm}\int_s^t \int_s^\tau \A(t)\A(\tau)\A(\tau')d\tau'd\tau\,\Theta+\cdots
 \end{align*}
 which converges provided $\|\A\|$ is finite over a time interval of interest. In addition,  the above series shows that $\star$-resolvents are nothing but time-ordered exponentials
 \begin{equation}\label{GreenOrder}
 \G=\mathcal{T}e^{\A(t)},
 \end{equation}
 while $\U=\Theta\star \G = \mathcal{T}e^{\int_s^t\A(\tau)d\tau}\,\Theta $.
 
  The advantage of the $\star$-formalism's outlook is that all methods and results from linear algebra are immediately applicable to the autonomous and non-autonomous differential settings in the distributional sense. This for example implies the existence of formal, exact and explicit $\star$-continued fraction representations of $\G$, known as path-sums \cite{Giscard2015, GisBon}. This also implies that standard procedure from linear algebra exist in the differential setting, such as Lanczos-triadiagonalization \cite{Giscard2020, Giscard2021, Giscard2023} and undoubtedly more.
 In this work we exploit the $\star$-product formalism to produce novel frame changes for ODEs and N-ODEs.

\section{Standard frame change}\label{EvalFrameStd}
As presented in Section~\ref{StdFrame}, the evolution operator $\U$ corresponding to matrix $\A(t)=\A_0(t)+\A_1(t)$ can be expressed as 
\begin{equation}
\U = \U_1 \star \left(\mathsf{I}_\star - \A_0\star \G_1\right)^{\star -1}.
\end{equation}
Let us now evaluate this explicitly. Firstly, concerning $\A_0\star \G_1$ we have
\begin{align}
\big(\A_0 \star \G_1\big)(t,s)&=\int_{-\infty}^{\infty}\A_0(t)\Theta(t-\tau)\G_1(\tau,s)d\tau,\nonumber\\
&=\A_0(t)\int_{-\infty}^{\infty}\!\Theta(t-\tau)\,\G_1(\tau,s)d\tau,\nonumber\\
&=\A_0(t)\,(\Theta\star \G_1)(t,s)= \A_0(t)\U_1(t,s),\label{SimplifyHG}
\end{align}
and since $t\geq s$ this is equivalent to $\A_0 \star \G_1=\A_0\U_1$. Since $\U_1(t,s)=\U_1(t)\U_1^{-1}(s)$ (and if $\A$ is Hermitian, $\U^{-1}=\U^\dagger$), it also follows that
\begin{align*}
(\A_0\star \G_1)^{\star n}&=\A_0(t) \U_1(t)\!\!\left(\int_s^t\!\!\!\U_1^{-1}(\tau) \A_0(\tau)\U_1(\tau)d\tau\!\right)^{\!\!\!\star n-1}\hspace{-4mm}\U_1^{-1}(s),
\end{align*}
so that,
\begin{align*}
&\left(\mathsf{I}_\star -\A_0\star \G_1\right)^{\star -1}(t,s)=\\
&\hspace{5mm}\mathsf{Id}_\star +\A_0(t)\U_1(t)\,\mathcal{T} e^{\int_s^t \U_1^{-1}(\tau)\A_0(\tau)\U_1(\tau)d\tau}\,\U_1^{-1}(s).
\end{align*}
Then
\begin{align*}
\U(t,s) &= \Big(\U_1\star\left(\mathsf{I}_\star -\A_0\star \G_1\right)^{\star -1}\Big)(t,s),\\
&=\U_1(t,s)+\U_1(t)\int_s^t \Big\{\U_1(\tau')^{-1}\A_0(\tau')\U_1(\tau')\\
&\hspace{20mm}\mathcal{T} e^{\int_s^{\tau'} \U_1^{-1}(\tau)\A_0(\tau)\U_1(\tau)d\tau}\,\U_1^{-1}(s)\Big\}d\tau',
\end{align*}
which evaluates to
\begin{align*}
\U(t,s)=\U_1(t)\,\mathcal{T}e^{ \int_s^t \U_1^{-1}(\tau)\A_0(\tau)\U_1(\tau)d\tau}\,\U_1^{-1}(s).
\end{align*}

 \section{Biframe presentation for $\dU$}\label{AppendixAlternative}
 
For $j=0,1$, by definition $\mathsf{G}_j := \delta' \star (\U_j) = \mathsf{I}_\star + \dU_j =(\mathsf{I}_\star - \A_j)^{\star -1}$ so we have $\dU_j = \A_j \star \G_j$.
Then, Eq.~(\ref{basicG}) yields
\begin{align*}
\mathsf{G} &= \G_0 \star \left(\mathsf{I}_\star-\dot{\mathsf{U}}_1\star \dot{\mathsf{U}}_0\right)^{\star-1} \star \G_1,\\
& = \G_0 \star \sum_{n=0}^\infty (\dot{\mathsf{U}}_1\star \dot{\mathsf{U}}_0)^{\star n} \star \G_1.
\end{align*}
The series on the second line is guaranteed to converge for the same reasons as Picard iterations: bounding $\|\dot{\u}_1\star \dot{\u}_0\|\leq C$, the norms of the $n$th $\star$-powers of $\dU_1\star\dU_0$ are $O(C^n/n!)$.
Given that $\dU=\G-\mathsf{I}_\star$, this result is equivalent to the following unconditionally convergent series representation for $\dU$ that reveals the symmetric roles played by $\dU_0$ and $\dU_1$,
\begin{align*}
\dU &=\dU_0+\dU_1+\dU_0\star \dU_1+\dU_1\star\dU_0\\
&\hspace{5mm}+\dU_0\star \dU_1\star \dU_0+\dU_1\star \dU_0\star \dU_1 \\
&\hspace{10mm}+\dU_0\star\dU_1\star \dU_0\star \dU_1+\dU_1\star\dU_0\star \dU_1\star \dU_0\\
&\hspace{15mm}+\cdots
\end{align*}
and from there $\mathsf{U}=\mathsf{I}\,\Theta+\Theta\star \dU$. The above series is just the sum over all alternating products of $\dU_0$ with $\dU_1$, manifestly invariant under exchange of indices $0\leftrightarrow 1$.

\section{Proof of Eq.~(\ref{BlueVersion})}\label{ProofBlue}
To arrive at Eq.~(\ref{BlueVersion}) from Eq.~(\ref{basicU}) we need only evaluating  
$\A_1\star \G_1\star \A_0 \star\G_0$. By Eq.~(\ref{SimplifyHG}) we have 
$
\A_j\star \G_j = \A_j(t)\U_j(t)\U_j^{-1}(s)
$
for $j=0,1$. Then 
 \begin{align*}
 \A_1\star \G_1\star \A_0 \star\G_0&=\int_s^t\A_1(t)\U_1(t)\U_1^{-1}(\sigma)\A_0(\sigma)\U_0(\sigma)\U^{-1}_0(s) d\sigma\\
 &=\A_1(t)\U_1(t)\!\int_s^t\!\U_1^{-1}(\sigma)\A_0(\sigma)\U_0(\sigma) d\sigma\,\U^{-1}_0(s).
 \end{align*}
It follows that the $\star$-resolvent of $\A_1\Theta\star \G_1\star \A_0\Theta \star\G_0$ is the time-ordered exponential of the above, which is Eq.~(\ref{BlueVersion}).

\section{Alternative form for the biframe}\label{ProofsDetail}
To obtain the biframe expression, instead of Eq.,~(\ref{RRRresult}) we can equally start from the ordinary identity between matrix resolvents 
$$
\mathsf{R}=\mathsf{R}_0.\big(\mathsf{I}-\mathsf{R}_1.\M_1.\mathsf{R}_0.\M_0\big)^{-1}.\mathsf{R}_1,
$$
which implies the alternative formula for Green's functions
\begin{equation}
\G=\G_0\star\big(\mathsf{I}_\star -  \G_1\star \A_1\Theta \star\G_0\star  \A_0\Theta\big)^{\star-1}\star \G_1.\label{BiSeriesRED}
\end{equation}
Now let us consider the content of the $\star$-resolvent in more details.
Using Eq.~(\ref{SimplifyHG}) we get
$$
\G_1\star \A_1\Theta \star\G_0\star  \A_0=\G_1\star(\A_1 \U_0\,\Theta)\star \A_0\Theta
$$
and so
\begin{align*}
&\big(\G_1\star \A_1 \star\G_0\star  \A_0\big)^{\star2},\\
&\hspace{5mm}=(\G_1\star(\A_1 \U_0)\star \A_0)^{\star2},\\
&\hspace{5mm}=\G_1\star(\A_1 \U_0)\star (\A_0 \U_1)\star(\A_1 \U_0)\star \A_0,
\end{align*}
since $(\A_0\star \G_1)(t,s)=\A_0(t)\U_1(t,s)$, again by Eq.~(\ref{SimplifyHG}). Continuing in this fashion  we verify by induction that for $n\geq 1$, 
\begin{align*}
&\big(\G_1\star \A_1 \star\G_0\star  \A_0\big)^{\star n}\\
&\hspace{5mm}=\G_1\star\big((\A_1 \U_0)\star (\A_0 \U_1)\big)^{\star n-1}\star(\A_1 \U_0)\star \A_0
\end{align*}
which yields, for $n\geq 1$,
\begin{align*}
&\left(\G_1\star \A_1 \star\G_0\star  \A_0\right)^{\star n}\star \G_1\\
&\hspace{3mm}=\G_1\star\big((\A_1 \U_0)\star (\A_0 \U_1)\big)^{\star n-1}\star(\A_1 \U_0)\star (\A_0\U_1)\\
&\hspace{3mm}=\G_1\star\big((\A_1 \U_0)\star (\A_0 \U_1)\big)^{\star n}.
\end{align*}
Finally, Eq.~(\ref{BiSeriesRED}) now reads
\begin{align*}
\G&=\G_0\star\G_1\sum_{n=0}^\infty \big((\A_1 \U_0)\star (\A_0 \U_1)\big)^{\star n},\\
&=\G_0\star \G_1\star \big(\mathsf{I}_\star - (\A_1 \U_0)\star (\A_0 \U_1)\big)^{\star-1},
\end{align*}
which immediately implies
\begin{equation}
\U=\U_0\star \G_1\star \big(\mathsf{I}_\star - (\A_1 \U_0)\star (\A_0 \U_1)\big)^{\star-1}.
\end{equation}
Since $\U_0(t,s)=\U_0(t)\U_0(s)$ and similarly for $\U_1$, we have
$$
(\A_1 \U_0)\star (\A_0 \U_1)=\A_1(t)\U_0(t)\int_s^t\U^{-1}_0(\tau)\A_0(\tau)\U_1(\tau)d\tau\,\U_1^{-1}(s).
$$
Putting everything together this leads to an alternative presentation of the biframe,
\begin{equation}
\U=\U_0\star \G_1\star \mathcal{T}e^{\A_1(t)\U_0(t)\int_s^t\U^{-1}_0(\tau)\A_0(\tau)\U_1(\tau)d\tau\,\U_1^{-1}(s)}\label{RedVersion}.
\end{equation}
Compare with Eq.~(\ref{BlueVersion}), in particular the content of the time-ordered exponential.
\section{Biframe change for constant Hamiltonians $\A$}\label{BiframeConstant}
In the situation where the coefficient matrix $\A$ is time-independent or, more generally, commutes with itself at different times, much simplifications occur with respect to the general formula provided in the main text. First of all, in such situation the biframe operator is invariant under time-translations $\mathcal{B}(t-s,0)=\mathcal{B}(t,s)$. This means that $\mathcal{B}$ effectively depends on a single variable (the difference $t-s$); and that all $\star$-products simplify to convolutions. Consequently the Laplace/Fourier domain may be employed, as for an autonomous ODE, and in this domain Eq.~(\ref{BlueVersion}) becomes
\begin{subequations}\label{LaplaceRel}
\begin{align}
\tilde{\mathcal{B}}(z)&=\tilde{\A}_1(z)\big(\mathsf{I}-\tilde{\A}_1(z)\big)^{-1}\tilde{\A}_0(z)\big(\mathsf{I}-\tilde{\A}_0(z)\big)^{-1},\label{LaplaceDom1}
\intertext{while the Laplace transform of the time-ordered exponential of $\mathcal{B}$ is $
\big(\mathsf{I}-\tilde{\mathcal{B}}(z)\big)^{-1}$ and so}
\tilde{\U}(z)&=z^{-1} \big(\mathsf{I}-\tilde{\A}_0(z)\big)^{-1}\big(\mathsf{I}-\tilde{\mathcal{B}}(z)\big)^{-1}\big(\mathsf{I}-\tilde{\A}_1(z)\big)^{-1}.\label{LaplaceDom2}
\end{align}
\end{subequations}
In these expressions $z$ is the Laplace/Fourier domain variable and $\tilde{\A}_i$, $\tilde{\mathcal{B}}$ are the Laplace/Fourier transforms of $\A_i$ and $\mathcal{B}$, respectively. Similarly $\tilde{\G}_1(z)=\big(\mathsf{I}-\tilde{\A}_1(z)\big)^{-1}$ and $\tilde{\U}_0(z)=z^{-1}\big(\mathsf{I}-\tilde{\A}_0(z)\big)^{-1}$.\

Although Eqs.~(\ref{LaplaceRel}) are only analytically correct  in the Laplace/Fourier domain when $\A$ self-commutes at different times, they are generally valid for numerical calculations in the time domain. Indeed, once time is discretized, $\star$-products become ordinary matrix products, and $\star$-resolvents turn into ordinary matrix resolvents; see \cite{DD21, GiscardForoozandeh, Pozza2024, PozzaBuggenhout2023, PozzaBuggenhout20232}. Then Eqs.~(\ref{BiFrame1}) directly take the form of Eqs.~(\ref{LaplaceRel}) with no Laplace/Fourier transformations involved, no matter how $\A(t)$ depends on $t$.

\end{document}